\documentstyle[12pt]{article}
\begin{document}
\begin{flushright}
hep-th/0309127\\
SNBNCBS-2003
\end{flushright}
\vskip 2cm
\begin{center}
{\bf \Large { Cohomological Operators and
Covariant Quantum Superalgebras }}

\vskip 3cm

{\bf R.P.Malik}
\footnote{ E-mail address: malik@boson.bose.res.in  }\\
{\it S. N. Bose National Centre for Basic Sciences,} \\
{\it Block-JD, Sector-III, Salt Lake, Calcutta-700 098, India} \\

\vskip 3cm

\end{center}

\noindent
{\bf Abstract}:
We obtain an interesting realization of the de Rham cohomological operators 
of differential geometry in terms of the noncommutative $q$-superoscillators 
for the supersymmetric quantum group $GL_{qp} (1|1)$. In particular, we show 
that a unique quantum superalgebra, obeyed by the bilinears of fermionic and 
bosonic noncommutative $q$-(super)oscillators of $GL_{qp} (1|1)$, is exactly 
identical to that obeyed by the de Rham cohomological operators. 
A set of discrete symmetry transformations for a set of 
$GL_{qp} (1|1)$ covariant quantum superalgebras turns out to be the analogue 
of the Hodge duality $*$ operation of differential geometry. A connection with 
an extended Becchi-Rouet-Stora-Tyutin (BRST) algebra obeyed by the conserved 
and nilpotent (anti-)BRST and (anti-)co-BRST charges, 
the conserved ghost charge 
and a conserved bosonic charge (which is equal to the anticommutator 
of (anti-)BRST and (anti-)co-BRST
charges) is also established. \\


\\$~~~~~~~~~~~~~~~~$noncommutative $q$-superoscillators; 
de Rham cohomological operators

\baselineskip=16pt


\newpage

\noindent
{\bf 1 Introduction}\\

\noindent
The subject of noncommutative geometry and corresponding noncommutative 
field theories has attracted a great deal
of interest during the past few years. Such an upsurge of
interest has been thriving because of its very clean and cogent appearance 
in the context of brane configurations related to the dynamics of string 
theories. In fact, the end points of the open strings, trapped on the D-branes,
turn out to be noncommutative 
\footnote{ It will be noted that, in the context of string theories and 
D-branes, it is the Snyder's idea of noncommutativity [1] that has become 
pertinent and popular.} 
in the presence of an antisymmetric ($B_{\mu\nu} = - B_{\nu\mu}$) potential 
that constitutes the 2-form (ie $B = \frac{1}{2}\;
(dx^\mu \wedge dx^\nu)\;B_{\mu\nu}$) background field for the whole 
system under consideration [2]. The
noncommutative supersymmetric gauge theories [3,4] are found to be
the low energy effective field theories for the $D$-branes 
discussed in the above. The experimental tests 
(eg, noncommutative Aharanov-Bohm effect; noncommutative 
synchrotron radiation, etc) for such kind 
of noncommutativity in the spacetime structure have been proposed [5,6] where 
it has been argued that
{\it only} the quantum mechanical effects are good enough to shed some
light on its very existence. This is why, in the recent past, a whole 
range of quantum mechanical studies has been performed for the 
noncommutative (non-)relativistic systems
and the ensuing results have gone into a systematic
understanding of this subject
from various physical and mathematical points of view
(see, eg, [7-10] and references therein).

The ideas behind the noncommutative (NC)
spacetime and spacetime quantization are
quite old ones (see, eg, [1,11,12] for details). A few decades ago, 
it was conjectured that the deformation of groups (ie 
the subject of quantum groups)
[13-17], based on the quasi-triangular Hopf algebras [18], together
with the idea of noncommutative geometry might shed some light on the 
existence of a ``fundamental length'' in the context of spacetime quantization.
It was also hoped
that this fundamental length will be responsible for 
getting rid of the infinities that plague the 
{\it local} quantum field theories (see, eg, [12] for more details). 
In our present investigation, we address some of the interesting 
issues associated with the noncommutativity present in the subject 
of quantum groups (without going into any kind of discussion on the Snyder's
idea of noncommutativity). 
It is worthwhile, in the context of quantum groups, to recall that some
interesting attempts have been made to construct the dynamics on a
NC quantum phase space by exploiting the differential geometry
and differential calculi developed on the noncommutative quantum hyperplanes
residing in the NC quantum cotangent manifolds
(see, eg, [19-22] and references therein). In particular, in [22], 
a consistent dynamics is constructed for the (non-)relativistic physical 
systems where a specific quantum group invariance and the 
ordinary (rotational) Lorentz invariance are respected together for any 
arbitrary ordering of the (space and) Lorentz spacetime indices. In a recent
paper [23], the noncommutativity due to the quantum groups and the 
noncommutativity due to the presence of a magnetic field in the 2D Landau
problem are brought together in the construction of a consistent
Hamiltonian and Lagrangian formulation where the symplectic 
structures, defined on the 4D cotangent manifold, play a very
important role. In this paper, a connection has
been attempted to be established between both kinds of noncommutativities 
(see, eg, [23] for details).
The $q$-deformed groups have also been treated as the gauge groups to
develop the $q$-deformed Yang-Mills theories which reduce to the 
ordinary Yang-Mills gauge theories in the limit $q \rightarrow 1$ 
(see, eg, [24,25] and references therein for details). In these endeavours,
the idea of quantum trace, quantum gauge orbits, quantum gauge
transformations, etc, have played notable roles [25,26].

The purpose of our present paper is to establish, in a single theoretical 
setting,  the inter-connections
among (i) the de Rham cohomological operators of differential geometry,
(ii) the $N = 2$ quantum mechanical superalgebra, and 
(iii) the extended BRST algebra for
some duality invariant gauge theories in the language of {\it noncommutative}
$q$-superoscillators for the supersymmetric quantum group $GL_{qp}(1|1)$.
We show that the bilinears of
the noncommutative $q$-superoscillators of the supersymmetric
quantum group $GL_{qp} (1|1)$ obey an algebra that is
reminiscent of the algebra obeyed by the de Rham cohomological
operators of differential geometry. It is also demonstrated that the
$GL_{qp} (1|1)$ covariant quantum superalgebras, obtained in our earlier
work [26], are {\it unique} and they reduce to a
{\it unique} superalgebra for the condition $pq = 1$. The latter
remains covariant, as is quite obvious, under the co-action of the 
supersymmetric quantum group $GL_{q, q^{-1}} (1|1)$ and the bilinears of the
$q$-superoscillators of this quantum group obey an algebra
that is reminiscent of the $N = 2$ supersymmetric quantum mechanics (SQM). 
At the SQM level too, an analogy with the de Rham cohomological operators 
is made, concentrating on the algebraic 
structure. The discrete symmetry transformations for all 
the three covariant quantum superalgebras turn out to be the analogue of the
Hodge duality $*$ operation of the differential geometry. This claim
has been shown at the level of the conserved 
and nilpotent charges corresponding to
the $N = 2$ supersymmetric quantum mechanical algebra (cf (6.10) and (6.11)
below) as well as
at the level of the conserved and nilpotent (co-)BRST charges 
(and corresponding nilpotent symmetry transformations) for the
duality invariant gauge theories (cf (6.4), (6.6) and (6.7) below)
and the corresponding
extended BRST algebra
\footnote{A concise discussion about the (co-)BRST symmetries, corresponding
nilpotent (co-)BRST charges and their extended BRST algebra, etc,
 for the duality invariant gauge theories,
has been given in section 6. Appropriate references on this
topic are cited at the beginning of this section.}. The identifications of the 
supercharges as well
as the (co-)BRST charges with the de Rham cohomological operators 
are in terms of the bilinears
of the noncommutative $q$-superoscillators of $GL_{qp} (1|1)$.

Besides the motivations pointed out in the above,
our present study is essential primarily on three counts. First, as is evident, 
the differential geometry and differential calculi 
play key roles in the discussion of a consistent dynamics in the framework
of Hamiltonian and/or Lagrangian formulation. Thus, it is an interesting 
endeavour to get some new noncommutative realization of the operators of the
differential geometry which might play important roles in the description
of the consistent noncommutative dynamics (see, eg, section 7 for more 
discussions). Second, to the best of our knowledge, the noncommutative 
realization of the cohomological operators, Hodge duality $*$ operation, Hodge
decomposition theorem, etc, has not been achieved so far in the language
of the quantum groups. It is, therefore, a challenging problem to obtain such
a realization. Finally, our present investigation {\it might} turn out to be
useful in the description of the $q$-deformed gauge theories where
the language of groups, differential geometry and differential-forms are 
exploited extensively. Furthermore, such studies might complement
(or provide an alternative to) the progress made in the realm of
NC gauge theories based on the Snyder's idea of noncommutativity.

The contents of our present investigation are organized as follows.
In section 2, we present a convenient synopsis of some
key concepts connected with the de Rham cohomological operators.
For our present paper to be self-contained, in
section 3, we recapitulate some preliminary results of our earlier work [26]
in a somewhat {\it different} manner. 
We derive a couple of covariant superalgebras for $GL_{qp}(1|1)$
in section 4. Sections 5 and 6 are central to our present 
paper. We deal with the discrete symmetries for the covariant 
superalgebras in section 5. These are shown to correspond to the Hodge 
duality $*$ operation of differential geometry in section 6. Furthermore,
in section 6, we also show the connection of some 
specific bilinears of the $q$-superoscillators of 
$GL_{qp}(1|1)$ and their $ N = 2$ SQM algebra with the 
BRST operators and their extended algebra. We make some concluding remarks 
in section 7 and point out a few future directions that 
could be pursued later.\\

\noindent
{\bf 2 Preliminary: de Rham cohomological operators}\\

\noindent
On a compact $D$-dimensional manifold without a boundary, there exist
three (ie $d, \delta, \Delta)$ cohomological operators in the realm
of differential geometry which are found to be responsible for the
study of the key and crucial properties associated with the differential
forms defined on the manifold. The (co-)exterior derivatives $(\delta)d$
are nilpotent  of order two (ie $ d^2 = \delta^2 = 0$) 
which could be readily 
proven by exploiting the basic definitions ($ d = dx^\mu \partial_\mu,
\delta = \pm * d *, dx^\mu \wedge dx^\nu = - dx^\nu \wedge dx^\mu$, etc)
of these operators in the inner product of the differential forms defined on
the $D$-dimensional compact manifold (ie $\mu = 0, 1, 2,.......D-1$). Here
$*$ is the Hodge duality operation defined on the manifold. The
Laplacian operator $\Delta = (d + \delta)^2 = d \delta + \delta d$
is defined 
in terms of the nilpotent (co-)exterior derivatives and is
a self-adjoint and a positive semi-definite quantity for a given 
compact manifold. The algebra obeyed by the above operators can be
succinctly expressed as 
$$
\begin{array}{lcl}
&& d^2 = 0 \;\qquad \delta^2 = 0\; \qquad \Delta = (d + \delta)^2 
= \{ d, \delta \} \nonumber\\
&& [\Delta, d ] = 0\; \qquad [\Delta, \delta ] = 0\; \qquad
\{ d, \delta \} \neq 0.
\end{array} \eqno(2.1)
$$
The above algebra shows that the Laplacian operator $\Delta$ is the Casimir
operator for the  whole algebra because it commutes with all
the cohomological operators [27-29].

The de Rham cohomology groups characterize the topology of a given
manifold in terms of the key properties associated with the
differential forms. These properties are, in a subtle way, captured by
the cohomological operators $d, \delta, \Delta$. In this context,
it is pertinent to point out that some of the properties that owe
their origin to the cohomological operators are (i) a
differential form $f_n$ of degree $n$ is said to be closed ($ d f_n = 0$)
and co-closed ($\delta f_n = 0$) if it is annihilated by $d$ and $\delta$,
respectively.
(ii) The same form is said to be exact ($ f_{n} = d e_{n-1}$) and
co-exact ($f_{n} = \delta c_{n +1}$) if and only if 
the above closed ($d f_{n} = 0$)
and co-closed ($\delta f_{n} = 0$) conditions are satisfied trivially due to
the nilpotentcy ($d^2 = \delta^2 = 0$) of the (co-)exterior derivatives 
$(\delta)d$. (iii) An $n$-form ($h_n$) is said to be a harmonic form if the 
Laplace equation $\Delta h_n = 0$ is satisfied which finally implies that the 
harmonic form $h_n$
is closed ($d h_n = 0$) and co-closed ($\delta h_n = 0$), simultaneously.
(iv) The celebrated Hodge decomposition theorem,
on a compact manifold without a boundary,  can be defined in terms of the
de Rham cohomological operators ($d, \delta, \Delta$) as (see, eg, [27-29]
for details)
$$
\begin{array}{lcl}
f_n = h_{n} \; + \; d e_{n-1} \; +\; \delta c_{n + 1}
\end{array} \eqno(2.2)
$$
which states that any arbitrary $n$-form $f_n$ (with
$ 0 \leq n \leq D$; $n = 0, 1, 2....$) on a $D$-dimensional compact 
manifold can be uniquely written as the sum of a harmonic form $h_n$, 
an exact form $d e_{n-1}$ and a co-exact form $\delta c_{n + 1}$. A close look
at (2.2) demonstrates that the degree of a form $f_{n}$
is raised by one if the exterior derivative $d$ acts on
it (ie $d f_{n} \sim g_{n+1}$). On the contrary, the degree of a form $f_{n}$
is lowered by one if it is acted
upon by the co-exterior derivative $\delta$ (ie $\delta f_{n} \sim g_{n-1}$). 
The degree of a form $f_{n}$ remains 
intact if it is acted upon by the Laplacian operator $\Delta$
(ie $\Delta f_{n} \sim g_{n}$).

Two closed ($ d f^\prime = d f = 0$) forms $f$ and $f^\prime$ are said
to belong to the same cohomology class with respect to the exterior
derivative $d$ if they differ by an exact form (ie $f^\prime = f + d g$,
for an appropriate non-zero form $g$).
Similarly, a co-cohomology can be defined w.r.t. $\delta$ where
any arbitrary two co-closed ($ \delta c^\prime = \delta c = 0$) forms 
$c^\prime$ and $c$ differ by a co-exact form (ie $ c^\prime = c + \delta m$,
for an appropriate non-zero form $m$). To wrap
up this section, we comment briefly on the $\pm$ signs present in the 
relationship ($\delta = \pm *\; d\; *$) between the (co-)exterior 
derivatives $(\delta)d$. By taking the inner product of the forms on the
$D$-dimensional manifold, it can be shown that, for an even $D$, we obtain
the relationship between $\delta$ and $d$ with a minus sign
(ie $ \delta = - * d *$). This conclusion 
is dictated by the fact that, in general, an
inner product between $n$-forms on the $D$-dimensional manifold leads to
the relationship between $\delta$ and $d$ as ( see, eg, [27] for details)
$$
\begin{array}{lcl}
\delta = (-1)^{(D n + D + 1)}\; *\; d\; *.
\end{array} \eqno(2.3)
$$
Thus, for an even dimensional manifold, there is always a minus sign on the 
r.h.s. and for the odd dimensional manifold, the above relation becomes:
$\delta = (-1)^{n}\;   *\; d\; *$ which shows that the $\pm$ signs 
on the r.h.s. for the 
latter case depend on the degree of the forms that are involved in the
specific inner product defined on the
odd dimensional manifold.\\

\noindent
{\bf 3 Quantum group $GL_{qp} (1|1)$ and $q$-superoscillators}\\

\noindent
In this section, we very briefly recapitulate,
in a somewhat different manner,  the bare essentials of our earlier work [26] 
which will be relevant for our further discussions. It can be seen that
the following transformations
$$
\begin{array}{lcl}
\left (\begin{array} {c} x \\ y \\
\end{array} \right )
\;\;\rightarrow \;\;
\left (\begin{array} {c}
x^\prime \\ y^\prime \\
\end{array} \right ) \;=\; 
\left (\begin{array} {cc}
a & \beta\\ \gamma & d\\
\end{array} \right )\;\;
\left (\begin{array} {c}
x \\ y \\
\end{array} \right ) \equiv (\;T\;) 
\left (\begin{array} {c}
x \\ y \\
\end{array} \right )  
\end{array}\eqno(3.1)
$$
$$
\begin{array}{lcl}
\left (\begin{array} {cc} \tilde x & \tilde y \\
\end{array} \right )
\;\;\rightarrow \;\;
\left (\begin{array} {cc}
\tilde x^\prime & \tilde y^\prime \\
\end{array} \right ) \;=\; 
\left (\begin{array} {cc} \tilde x & \tilde y \\
\end{array} \right )\;
\left (\begin{array} {cc}
a & \beta\\ \gamma & d\\
\end{array} \right ) \equiv 
\left (\begin{array} {cc} \tilde x & \tilde y \\
\end{array} \right )\; (\; T\;)
\end{array}\eqno(3.2)
$$
for the even pair of variables $(x, \tilde x)$ and odd 
($y^2 = \tilde y^2 = 0)$ pair of
variables $(y, \tilde y)$, that define the super quantum hyperplane, with
conditions
$$
\begin{array}{lcl}
&& x\; y = q \;y\; x \rightarrow x^\prime\; y^\prime 
= q\; y^\prime\; x^\prime\; \qquad
 y^2 = 0 \;\;\rightarrow \;\;
(y^\prime)^2 =  0 \nonumber\\
&& \tilde x \;\tilde y = p\; \tilde y \;\tilde x \rightarrow
\tilde x^\prime\; \tilde y^\prime = p \;\tilde y^\prime\; \tilde x^\prime\; 
\qquad \tilde y^2 = 0 \;\;\rightarrow \;\;
(\tilde y^\prime)^2 = 0
\end{array} \eqno(3.3)
$$
lead to the braiding relationships among the rows and columns,
constituted by the even elements $(a, d)$ and odd $(\beta^2 = \gamma^2 = 0)$
elements $(\beta, \gamma)$ of the $ 2\times 2$ supersymmetric 
quantum matrix $T$ defined in (3.1) and (3.2), as
$$
\begin{array}{lcl}
&& a \beta = p \beta a\;\; \qquad \;a \gamma = q \gamma a\;\; 
\qquad \;\;\beta \gamma= - \;(q/p)\; \gamma \beta\; \;\qquad\;\;\;
d \gamma = q \gamma d \nonumber\\
&& d \beta = p \beta d\; \quad \beta^2 = \gamma^2 = 0\; \quad a d - d a
= - (p - q^{-1}) \beta \gamma = (q - p^{-1}) \gamma \beta.
\end{array}\eqno(3.4)
$$
In fact, this method is the simplified version of
the Manin's super quantum hyperplane approach to the construction of
the general quantum super groups [17]. For $q = p$, the above relations
boil down to the braiding relations in the rows and columns for the
supersymmetric quantum group $GL_{q} (1|1)$ with a single deformation
parameter $q$. For the special case of $ pq = 1$, which corresponds to the
supersymmetric quantum group $GL_{q,q^{-1}}(1|1)$, we obtain the following
relations among the elements of $GL_{q,q^{-1}} (1|1)$ (that emerge from (3.4)):
$$
\begin{array}{lcl}
&& a \beta = q^{-1} \beta a\; \quad \;a \gamma = q \gamma a\;\; 
\quad \;\;\beta \gamma= - \;(q^2)\; \gamma \beta\; \quad\;
d \gamma = q \gamma d \nonumber\\
&& d \beta = q^{-1} \beta d\; \;\;\qquad \;\beta^2 = 0\; \;\;\qquad\;
\gamma^2 = 0\; \;\;\;\qquad \;\;a d = d a.
\end{array}\eqno(3.5)
$$
The above relations (3.4) and (3.5)
will turn out to be quite useful for our later discussions. To study
the $GL_{qp} (1|1)$ covariant relations among the $q$-superoscillators,
we introduce a pair of 
noncommutative bosonic oscillators $(A, \tilde A)$ and a pair
of noncommutative fermionic (ie $B^2 = \tilde B^2 = 0$) 
oscillators $(B, \tilde B)$.
It is straightforward to check that the following $GL_{qp} (1|1)$
transformations of the super column matrix $(A, B)^T$ and 
the super row matrix ($\tilde A, \tilde B)$ constructed by the 
superoscillators
$$
\begin{array}{lcl}
\left (\begin{array} {c} A \\ B \\
\end{array} \right )
\;\;\rightarrow \;\;
\left (\begin{array} {c}
A^\prime \\ B^\prime \\
\end{array} \right ) \;=\; 
\left (\begin{array} {cc}
a & \beta\\ \gamma & d\\
\end{array} \right )\;\;
\left (\begin{array} {c}
A \\ B \\
\end{array} \right ) \equiv (\;T\;) 
\left (\begin{array} {c}
A \\ B \\
\end{array} \right )  
\end{array}\eqno(3.6)
$$
$$
\begin{array}{lcl}
\left (\begin{array} {cc} \tilde A & \tilde B \\
\end{array} \right )
\rightarrow 
\left (\begin{array} {cc}
\tilde A^\prime & \tilde B^\prime \\
\end{array} \right ) &=&
\left (\begin{array} {cc} \tilde A & \tilde B \\
\end{array} \right )
\left (\begin{array} {cc}
a^{-1} (1 + \beta d^{-1} \gamma a^{-1}) & - a^{-1} \beta d^{-1}\\ 
- d^{-1} \gamma a^{-1} & d^{-1} (1 - \beta a^{-1} \gamma d^{-1})\\
\end{array} \right) \nonumber\\
&\equiv& 
\left (\begin{array} {cc} \tilde A & \tilde B \\
\end{array} \right )\; (\; T\;)^{-1}
\end{array}\eqno(3.7)
$$
leave the following algebraic relationships invariant
$$
\begin{array}{lcl}
 A B = q\; B A\; \qquad \tilde B \tilde A = p \;\tilde A \tilde B\;
\qquad B^2 = \tilde B^2 = 0
\end{array} \eqno(3.8)
$$
if we exploit the $q$-commutation relations of (3.4). Consistent with (3.8),
the other general covariant relations among the superoscillators are
$$
\begin{array}{lcl}
&& A \tilde B  = {\displaystyle \frac{(\lambda - \nu)} {q}} \tilde B A\;
\;\qquad B \tilde A   = {\displaystyle \frac{(\lambda - \nu)} {p}} \tilde A B 
\nonumber\\
&& A \tilde A - {\displaystyle \frac{(\lambda - \nu)}{pq}} \tilde A A 
= B \tilde B + {\displaystyle \frac{(\lambda - \nu)}{pq}}
\; \tilde B B 
\end{array}\eqno(3.9)
$$
if we assume the validity of the following general relation between the
bilinears constructed from the bosonic oscillators $A, \tilde A$
as well as the fermionic oscillators $B, \tilde B$
\footnote{It should be emphasized that, in our earlier work [26], we have
postulated the validity of a different kind of relationship
(ie $B \tilde B + \nu \tilde B B = 1 + \lambda \tilde A A$) among
the bilinears of the bosonic and fermionic (super)oscillators.}
$$
\begin{array}{lcl}
 A \tilde A - \lambda \tilde A A = 1 + \nu \tilde B B 
\end{array}\eqno(3.10)
$$
where $\lambda$ and $\nu$ are some arbitrary non-zero commuting parameters
which can be determined by exploiting the {\it associativity} of the trilinear 
combinations of the (super)oscillators. 
This associativity requirement is, in fact, equivalent to invoking 
the sanctity of the graded Yang-Baxter equations {\it vis-{\`a}-vis} the 
covariant algebraic relations (cf (3.8)-(3.10)). 
It will be noted that all the relations in (3.9) actually emerge from
(3.10) when we exploit the basic transformations (3.6) and (3.7)
on the (super)oscillators and use the relations in (3.4) and (3.8).
As a side remark, we wish to state that the last
algebraic superoscillator relation of (3.9) and our assumption (3.10) imply
the following relationship between the bilinears of the (super)oscillators
$$
\begin{array}{lcl}
B \tilde B = 1 
+ \Bigl ( \lambda - {\displaystyle \frac{\lambda - \nu}{pq}} \Bigr ) 
\;\tilde A  A + \Bigl (\nu - {\displaystyle \frac{\lambda - \nu}{pq}} 
\Bigr ) \; \tilde B B
\end{array} \eqno(3.11)
$$
where the r.h.s. contains terms with all the tilde oscillators arranged towards
the left and all the untilde oscillators arranged towards the right. This
relationship will turn out to be quite helpful 
in the next section where we shall
invoke the associativity condition. Taking into account the explicit
transformations in (3.6), (3.7) and relations in (3.4), (3.8), it can be 
checked that all the (super)oscillator
relations from (3.8) to (3.11) are {\it covariant}
under the co-action of supersymmetric quantum group $GL_{qp} (1|1)$.\\

\noindent
{\bf 4 Covariant quantum superalgebras for $GL_{qp} (1|1)$}\\

\noindent
In this section, we establish the fact that a set
of a couple of covariant superalgebras,
obtained in our earlier work [26], is a {\it unique} set of algebras
for the quantum group $GL_{qp}(1|1)$. To this end in mind,
we compute the exact values of the parameters $\lambda$ and $\nu$ in the above
from the requirement that in the set of, for instance, a trilinear
(super)oscillators $B \tilde B \tilde A$, we
can bring all the tilde operators to the left in two different 
ways as listed below
$$
\begin{array}{lcl}
(B\tilde B) \;\tilde A &=& 
\Bigl [\;1 + \lambda - 
{\displaystyle \frac{\lambda - \nu}{pq}}
\;\Bigr ]\; \tilde A + \lambda\; \Bigl [ \;\lambda - 
{\displaystyle \frac{\lambda - \nu}{pq}} \;\Bigr ] \;
\tilde A \tilde A A 
\nonumber\\
& +& \Bigl [\; \nu \Bigl ( \lambda -
{\displaystyle \frac{\lambda - \nu}{pq}} \Bigr ) \; 
+ (\lambda - \nu) \Bigl ( \nu -
{\displaystyle \frac{\lambda - \nu}{pq}} \Bigr ) \;\Bigr ]\;
\tilde A \tilde B B
\end{array}\eqno(4.1) 
$$
$$
\begin{array}{lcl}
B \;(\tilde B \tilde A) &=& 
\Bigl [\; \bigl (\lambda - \nu \bigr )
\;\Bigr ]\; \tilde A + \bigl (\lambda - \nu
\bigr ) \; \Bigl [\; \lambda - 
{\displaystyle \frac{\lambda - \nu}{pq}} \;\Bigr ] \;
\tilde A \tilde A A \nonumber\\
& + & \Bigl [\;(\lambda - \nu) \Bigl ( \nu -
{\displaystyle \frac{\lambda - \nu}{pq}} \Bigr ) \;\Bigr ]\;
\tilde A \tilde B B.
\end{array}\eqno(4.2) 
$$
At this crucial juncture,
a couple of remarks are in order. First,
it will be noted that in (4.1) as well as (4.2), we have chosen a different
set of (super)oscillators than the one chosen in our earlier work 
\footnote {For the proof of associativity, a trilinear set $A\tilde A\tilde B$ 
has been chosen in [26] for the purpose of reordering it in two different ways.
More such kind of trilinear combinations of (super)oscillators
can be considered for the determination of $\nu$ and $\lambda$
(see, e.g., Appendix for details). However, the 
covariant quantum superalgebras (ie (3.8),(4.4),(4.5))
remain the same. This demonstrates clearly the {\it uniqueness} of these basic 
superalgebras that are present in (3.8), (4.4), (4.5) as well as in their
special case (4.6).} [26]. Second, the 
expressions for the (super)oscillators on the r.h.s. of the equations
(4.1) and (4.2) are unique as far as all the covariant algebraic relations
among the (super)oscillators from (3.8) to (3.11) are concerned.
For the validity of the associativity condition, it is essential
that the r.h.s. of both the above equations should match with each-other.
Such an equality imposes the following two conditions on $\lambda$ and $\nu$
$$
\begin{array}{lcl}
(i) \;\;\;\;\nu = 0 \qquad \lambda = p q\; \;\;\;\qquad
(ii)\;\;\;\; \nu = {\displaystyle \frac{( 1 - p q)}{pq}}\; \qquad \lambda = 
{\displaystyle \frac{1}{pq}}.
\end{array} \eqno(4.3)
$$
It should be re-emphasized that the above 
{\it associativity} requirement is equivalent to
the validity of the graded Yang-Baxter equation in the context of 
supersymmetric
quantum group $GL_{qp} (1|1)$. The covariant superalgebra for the
bilinears corresponding to the case (i) are
$$
\begin{array}{lcl}
&&B \tilde A = q \tilde A B\; \qquad A \tilde B = p \tilde B A\; \qquad
A \tilde A - pq\; \tilde A A = 1\; \nonumber\\
&& B \tilde B\; + \;\tilde B B
= 1\; + \;(pq - 1)\; \tilde A A\;
\end{array} \eqno(4.4)
$$
which are, in addition to the invariant relations (3.8) for the bilinears.
For the case (ii), in addition to (3.8), 
the other bilinear $q$-superoscillator relations are
$$
\begin{array}{lcl}
&&B \tilde A = p^{-1}\; \tilde A B \qquad A \tilde B = q^{-1} 
\tilde B A\; \qquad
B \tilde B + \tilde B B = 1\; \nonumber\\
&& A \tilde A\; - \;{\displaystyle \frac{1}{pq}}\; \tilde A A
= 1\; + \;{\displaystyle \frac{(1 - pq)}{pq}} \;\tilde B B.
\end{array} \eqno(4.5)
$$
It should be emphasized, at this stage, that the relations
(3.8), (4.4) and (4.5) are same as the ones obtained in our earlier work [26]
where a different set of superoscillators 
(in the trilinear form) was taken into consideration. 
For the case when $ pq = 1$, we obtain a {\it unique} 
solution ($\nu = 0, \; \lambda = 1$) where the algebraic relations
(3.8), (4.4) and (4.5) reduce to 
$$
\begin{array}{lcl}
&&B \tilde A = q \tilde A B\; \qquad A \tilde B = q^{-1}\; \tilde B A\; \qquad
A \tilde A - \tilde A A = 1\; \qquad
B^2 = 0\; \nonumber\\
&& B \tilde B + \tilde B B= 1\; \qquad A B = q B A\; \qquad
\tilde B \tilde A = q^{-1} \tilde A \tilde B\; \qquad
\tilde B^2 = 0.
\end{array} \eqno(4.6)
$$
From the $q$-superoscillators $(A, \tilde A, B, \tilde B)$,
 one can construct a four-dimensional
``adjoint representation'' for the supersymmetric quantum group 
$GL_{qp} (1|1)$ in terms of the following four bilinears [26]
$$
\begin{array}{lcl}
Y = {\displaystyle \frac{A \tilde A + \mu B \tilde B}{1 + \mu}}\; \quad
H = A \tilde A - B \tilde B\; \quad Q = A \tilde B\; \quad \bar Q = B \tilde A
\end{array} \eqno(4.7)
$$
where $\mu \neq -1$ and the specific operator $H = A \tilde A - B \tilde B$
is invariant under the co-action of the supersymmetric quantum group
$GL_{qp} (1|1)$. It is worthwhile to emphasize that the above operator
$H$ has been derived in [26] 
by exploiting the idea of supertrace for a $2 \times 2$
super quantum matrix constructed from the $q$-superoscillators
$A, \tilde A, B, \tilde B$. The operators in (4.7) obey the following
superalgebra which turns out to be the
reminiscent of the $N = 2$ supersymmetric quantum algebra:
$$
\begin{array}{lcl}
&& [H, Q] = [H, \bar Q] = [H, Y] = 0\; \qquad Q^2 = \bar Q^2 = 0\; \nonumber\\
&& \{Q, \bar Q\} = H\; \qquad [Q, Y] = + Q\; \qquad [\bar Q, Y] = - \bar Q.
\end{array} \eqno(4.8)
$$
The above supersymmetric algebra is true for the case when
$\nu = (pq)^{-1} (1 - pq), \lambda = (pq)^{-1}$ 
(ie the case (ii) in eqn. (4.3)) as well as for the case
when $pq = 1$. In the latter case, both the conditions of (4.3) reduce to 
a single condition (ie
$\nu = 0, \lambda = 1$). Such kind of algebra for the case when
$\nu = 0, \lambda = pq$ (ie the case (i) of (4.3))
is as follows
$$
\begin{array}{lcl}
&& [H, Q] = 0\; \qquad [H, \bar Q] = 0\; \qquad [H, Y] = 0\; 
\qquad Q^2 = \frac{1}{2}\; \{ Q, Q\} = 0\; \nonumber\\
&& \{Q, \bar Q\} = \bigl [\;1 + (pq -1)\; H \;\bigr ]\; H
\qquad \bar Q^2 = 
\frac{1}{2}\; \{ \bar Q, \bar Q \} = 0\; \nonumber\\
&& [Q, Y] = + \;\bigl [\; 1 + (pq -1)\; H \;\bigr ]\; Q\; \qquad
 [\bar Q, Y] = -\; \bigl [\; 1 + (pq -1)\; H\; \bigr ]\; \bar Q.
\end{array} \eqno(4.9)
$$
Even though (4.9) looks a bit different from (4.8), it can be seen that
the following redefinitions of $Y, Q, \bar Q$ in terms of 
$\hat Y, P, \bar P$
$$
\begin{array}{lcl}
\hat Y = {\displaystyle \frac{Y}{1 + (pq - 1)\; H}}\; \quad
P = {\displaystyle \frac{Q}{[1 + (pq - 1)\; H]^{(1/2)}}}\; \quad
\bar P = {\displaystyle \frac{\bar Q}{[1 + (pq - 1)\; H]^{(1/2)}}}\; 
\end{array} \eqno(4.10)
$$
lead to the $N =2$ supersymmetric quantum mechanical superalgebra
$$
\begin{array}{lcl}
&& [H, P] = [H, \bar P] = [H, \hat Y] = 0\; \qquad P^2 = \bar P^2 = 0\; 
\nonumber\\
&& \{P, \bar P\} = H\; \qquad [P, \hat Y] = + P\;
 \qquad [\bar P, \hat Y] = 
- \bar P.
\end{array} \eqno(4.11)
$$
The identification in (4.10) is valid for derivation of (4.11) because
$H$ is the Casimir operator for (4.9) and it does commute with the
original operators $Q, \bar Q$ and $Y$.
It is crystal clear that the four operators in (4.7) do give a realization
of $N = 2$ supersymmetric quantum mechanics in terms of the 
{\it noncommutative}
$q$-superoscillators. Here the Hamiltonian $H$ is invariant under
$GL_{qp} (1|1)$ transformations (3.6) and (3.7),
$Q$ and $\bar Q$ are like nilpotent supercharges
and $Y$ is like Witten index which encodes the fermion number
for the supersymmetric  quantum mechanical theory.

At this stage, we summarize the main results of our present section.
First, the superoscillator algebraic relations (3.8) remain {\it invariant}
under the co-action of the supersymmetric quantum group $GL_{qp} (1|1)$.
Second, the requirement of associativity condition leads to only two
$GL_{qp} (1|1)$ {\it covariant} algebraic relations (cf (4.4) and (4.5)
in addition to (3.8))
for the specific
values of $\lambda$ and $\nu$ as given in (4.3). Third, the above
two {\it covariant} relations reduce to a unique algebraic relation (4.6)
which is found to be $GL_{q, q^{-1}} (1|1)$ covariant under the co-action
of $GL_{qp} (1|1)$ for the deformation parameters satisfying $pq = 1$.
Fourth, this unique superalgebra provides a unique realization of the
de Rham cohomology operators of differential geometry as there is one-to-one
correspondence between bilinears of the $q$-superoscillators
and the cohomological operators (ie $Q \rightarrow d, \bar Q 
\rightarrow \delta, H \rightarrow \Delta$) as can be seen in (4.8). The 
operator $Y$ is the analogue of the Witten index which determines the 
{\it degree of the forms} in the language of fermion numbers 
\footnote{For the supersymmetric quantum mechanical theory, it can be
checked that one can choose
$Q = \sigma_{+} = \frac{1}{2} (\sigma_1 + i \sigma_2), \bar Q = \sigma_{-}
= \frac{1}{2} (\sigma_{1} - i \sigma_2), Y = \frac{1}{2} (1 - \sigma_3)$ 
in terms of the $ 2 \times 2$ Pauli matrices which do satisfy
$ [ Q, Y ] = + Q, [\bar Q,  Y] = - \bar Q, Q^2 = \bar Q^2 = 0$
(see, eg, [30] for further references and more details).}
of the supersymmetric theory
(see, eg, [31] for details). Fifth, the analogue of
the Hodge duality $*$ operation of differential geometry
turns out to be a host of discrete symmetry transformations
for the superalgebras (3.8), (4.4), (4.5) and (4.6)
which are discussed in section 6 (see below).\\

\noindent
{\bf 5 Discrete symmetries for the covariant superalgebras}\\

\noindent
It is interesting to note that the superalgebra (4.6) for $pq = 1$ 
(ie $\nu = 0, \; \lambda = 1$) remains form-invariant under the following
discrete symmetry transformations
$$
\begin{array}{lcl}
A \rightarrow \pm i \tilde A\; \quad \tilde A \rightarrow \pm i A\; \quad 
B \rightarrow \pm \tilde B\; \quad \tilde B \rightarrow \pm B.
\end{array} \eqno(5.1)
$$
Notice that there is no transformation on the deformation parameters
$p$ and $q$ but there are transformations on the superoscillators
$A, \tilde A, B , \tilde B$. Similar kind of a couple of 
discrete symmetry transformations for 
$\nu = 0, \lambda = pq$ in the case of covariant superalgebra 
(4.4) (together with the relations (3.8)) are
$$
\begin{array}{lcl}
A \rightarrow \pm i q \tilde A\; \quad \tilde A \rightarrow \pm i p A\; \quad 
B \rightarrow \pm q \tilde B\; \quad \tilde B \rightarrow \pm p B\;
\quad q \rightarrow p^{-1}\; \quad p \rightarrow q^{-1}
\end{array} \eqno(5.2a)
$$
$$
\begin{array}{lcl}
A \rightarrow \pm i p \tilde A\; \quad \tilde A \rightarrow \pm i q A\; \quad 
B \rightarrow \pm p \tilde B\; \quad \tilde B \rightarrow \pm q B\;
\quad q \rightarrow p^{-1}\; \quad p \rightarrow q^{-1}.
\end{array} \eqno(5.2b)
$$
The covariant superalgebra (4.5) (together with (3.8))
corresponding to $\nu = (pq)^{-1}
(1 - pq), \lambda = (pq)^{-1}$ is found to be endowed with the following
discrete symmetry transformations on the superoscillators $(A, \tilde A,
B, \tilde B)$ and the deformation parameters $p$ and $q$:
$$
\begin{array}{lcl}
A \rightarrow \pm i  \tilde A\; \quad \tilde A \rightarrow \pm i  A\; \quad 
B \rightarrow \pm \tilde B\; \quad \tilde B \rightarrow \pm  B\;
\quad q \rightarrow p^{-1}\; \quad p \rightarrow q^{-1}.
\end{array} \eqno(5.3)
$$
It can be checked that, under the transformations (5.1) and (5.3), the
four bilinear operators of (4.7) individually undergo the following change
$$
\begin{array}{lcl}
&& H = A \tilde A - B \tilde B\; \;\;\rightarrow\;\;\; \tilde H = - \tilde A A
- \tilde B B\; \nonumber\\
&& Q = A \tilde B \;\rightarrow \;\tilde Q = \pm \; i \tilde A B\; \qquad
\bar Q = B \tilde A \;\rightarrow \;\tilde {\bar Q} = \pm \;i \tilde B A\; 
\nonumber\\
&& Y = {\displaystyle \frac{A \tilde A + \mu B \tilde B}{1 + \mu}}
\;\;\;\rightarrow \;\;\;
\tilde Y = {\displaystyle \frac{- \tilde A A + \mu \tilde B B}{1 + \mu}}.
\end{array} \eqno(5.4)
$$
It is elementary to check that the $N = 2$ supersymmetric 
quantum algebra (4.8),
for the bilinears in (4.7), remains {\it form} 
invariant under (5.4). This can be succinctly stated in the mathematical form
as listed below
$$
\begin{array}{lcl}
&& [\tilde H, \tilde Q] = [\tilde H, \tilde {\bar Q}] 
= [\tilde H, \tilde Y] = 0\; \qquad \tilde Q^2 = (\tilde {\bar Q})^2 = 0\; 
\nonumber\\
&& \{\tilde Q, \tilde {\bar Q}\} = \tilde H\; \qquad 
[\tilde Q, \tilde Y] = + \tilde Q\; \qquad [\tilde {\bar Q}, \tilde Y ] = 
- \tilde {\bar Q}.
\end{array} \eqno(5.5)
$$
Now let us concentrate on the discrete transformations in (5.2). It
is straightforward to see that the four bilinears of (4.7) transform
in the following manner under (5.2)
$$
\begin{array}{lcl}
&& H = A \tilde A - B \tilde B\; \;\;\rightarrow\;\;\; \tilde H = - 
(pq)\; (\tilde A A + \tilde B B)\; \nonumber\\
&& Q = A \tilde B \;\rightarrow \;\tilde Q = \pm \; i (pq)\; 
(\tilde A B)\; \qquad
\bar Q = B \tilde A \;\rightarrow \;\tilde {\bar Q} = \pm \;i (pq)\;
 (\tilde B A)\; \nonumber\\
&& Y = {\displaystyle \frac{A \tilde A + \mu B \tilde B}{1 + \mu}}
\;\;\;\rightarrow \;\;\;
\tilde Y = - (pq) \;
\Bigl [\; {\displaystyle \frac{ \tilde A A - \mu \tilde B B}{1 + \mu}}
\;\Bigr ].
\end{array} \eqno(5.6)
$$
These transformed operators obey the following algebra
$$
\begin{array}{lcl}
&& [\tilde H, \tilde Q] = 0\; \qquad [\tilde H, \tilde {\bar Q}] = 0\; 
\qquad [\tilde H, \tilde Y] = 0\; 
\qquad (\tilde Q)^2 = \frac{1}{2}\; \{ \tilde Q, \tilde Q\} = 0\; \nonumber\\
&& \{\tilde Q, \tilde {\bar Q}\} = \bigl [1 \;
+ (\frac{1}{pq} -1)\; \tilde H \;\bigr ]\;  \tilde H\; 
\qquad (\tilde {\bar Q})^2 = 
\frac{1}{2}\; \{ \tilde {\bar Q}, \tilde {\bar Q} \} = 0\; \nonumber\\
&& [\tilde Q, \tilde Y] = + \bigl [\; 1 
+ (\frac{1}{pq} -1)\; \tilde H \;\bigr ]\; \tilde Q\; \qquad
 [\tilde {\bar Q}, \tilde Y ] = - \bigl [\; 1 
+ (\frac{1}{pq} -1)\; \tilde  H \;\bigr ]\; \tilde {\bar Q}.
\end{array} \eqno(5.7)
$$
As far as the superalgebras in (4.8), (4.9), (5.5) and (5.7) are concerned,
there are a few comments in order. First, it can be noted that 
the deformation parameters do not appear in the algebra (4.8) for the
cases (i) $\nu = 0, \lambda = 1$ (ie the case when $pq = 1$), and 
(ii) $\nu = (pq)^{-1} (1 - pq), \lambda = (pq)^{-1}$. As a result,
the corresponding algebra (5.5) for the transformed bilinears
remain form-invariant. Second, in the
case of $\nu = 0, \lambda = pq$, the superalgebra (4.9) contains deformation
parameters for the bilinears of (4.7). This is why the corresponding 
superalgebra (5.7) for the tilde operators contains a $(pq)^{-1}$
at the place of $pq$ occurring in (4.9). The latter is due to the fact
that, in the discrete transformations (5.2), we have $p \rightarrow q^{-1},
q \rightarrow p^{-1}$. Third, it is very interesting to point out that,
for the restriction $pq = 1$, the discrete symmetry transformations
in (5.3) do converge trivially to (5.1). Fourth, let us concentrate on (5.2a)
which leads to: $A \rightarrow i q \tilde A, \tilde A \rightarrow i q^{-1} A,
B \rightarrow q \tilde B, \tilde B \rightarrow q^{-1} B$ if we choose the
upper signs. It can be readily checked that the algebraic relations (4.6)
remain invariant under above discrete transformations too. The same can be
checked to be true for the other transformations in (5.2a) and (5.2b) as well.
The key point to be noted is that these transformations 
(cf (5.2a) and (5.2b)) owe their origin
to (5.1) (ie for $pq = 1$) when one plays with some constant factors 
(eg, $q$ and $q^{-1}$) that are plugged in the transformations (5.1). 
Fifth, the discrete symmetry transformations
in (5.1), (5.2) and (5.3) would turn out to be the analogue of Hodge
duality $*$ operation of differential geometry as we shall see in the
next section.  In fact, we shall establish this analogy at the level
of a duality between supercharges $Q$ and $\bar Q$ {\it themselves}
as well as at the
level of symmetry transformations generated by $Q$ and $\bar Q$ which
are cast in the language of BRST and co-BRST symmetries, respectively.\\

\noindent
{\bf 6 Connection with the extended BRST algebra}\\

\noindent
In a recent set of papers (see, eg, [32-39]), a connection between the
de Rham cohomology operators $(d, \delta, \Delta)$ and (anti-)BRST
charges $Q_{(a)b}$, (anti-)co-BRST charges $Q_{(a)d}$, a ghost charge $Q_g$
and a bosonic charge $W = \{Q_b, Q_d \} = \{ Q_{ab}, Q_{ad} \}$ has been 
established for (i) the free Abelian 1-form gauge theory [31-33],
(ii) the self-interacting 1-form non-Abelian gauge theory (where there is no
interaction between the matter fields and gauge field) [34,35], (iii) the
interacting 1-form $U(1)$ gauge theory where there is an interaction between
1-form Abelian gauge field and the matter (Dirac) fields [36,37], and
(iv) the free Abelian 2-form gauge theory [38,39], etc, in the language
of symmetry properties for the Lagrangian density of these theories. In all 
the above examples of the field theoretic models, the algebra satisfied by
the local and conserved charges are found to be
$$
\begin{array}{lcl}
&& [W, Q_{r} ] = 0\; \qquad r = g, b, ab, d, ad\; \qquad
Q_{b}^2 = Q_{d}^2 = Q_{ab}^2 = Q_{ad}^2 = 0\; \nonumber\\
&& W = \{Q_{b}, Q_{d} \} = \{ Q_{ab}, Q_{ad} \}\; \quad
\{ Q_{b}, Q_{ad} \} = \{Q_{d}, Q_{ab} \} = 0\; \quad
\{ Q_{b}, Q_{ab} \} = 0 \nonumber\\
&& i [ Q_{g}, Q_{b(ad)} ] = + Q_{b(ad)}\; \qquad
i [ Q_{g}, Q_{d(ab)} ] = - Q_{d(ab)}\; \qquad
\{ Q_{d}, Q_{ad} \} = 0.
\end{array} \eqno(6.1)
$$
The above algebra is exactly like the algebra obeyed by the de Rham
cohomological operators with a two-to-one mapping
between the conserved charges and cohomological operators: $Q_{b(ad)}
\rightarrow d, \; Q_{d(ab)} \rightarrow \delta, \; W = \{Q_{b}, Q_{d} \}
= \{ Q_{ab}, Q_{ad} \} \rightarrow \Delta$. For all the above models,
a set of discrete symmetry 
transformations has been shown to correspond to the Hodge duality
$*$ operation of differential geometry. Furthermore, the analogue of
the Hodge decomposition theorem (2.2) has been derived in the quantum
Hilbert space of states where the {\it ghost number} plays the role of the
{\it degree} of the differential forms. Thus, above examples provide a 
beautiful
set of field theoretical models for the Hodge theory where all the 
cohomological operators, Hodge duality $*$ operation, Hodge decomposition
theorem, etc, are expressed in terms of the local,
covariant  and continuous (as well as discrete) symmetry
transformations and their corresponding generators (ie conserved charges).

Now we shall concentrate on the {\it unique} algebra (4.8) which is separately
valid for (i) $\nu = (pq)^{-1} (1 - pq), \lambda = (pq)^{-1}$, and
(ii) $ \nu = 0, \lambda = 1$ (ie the case when $pq = 1$). In fact, as 
emphasized earlier, both the covariant quantum superalgebras (4.8) and (4.9)
do converge to (4.8) for the case $pq = 1$. In contrast to the ``two-to-one''
mapping between local and conserved charges and the cohomological operators
for the algebra (6.1), we shall see that the algebra (4.8) provides a 
``one-to-one'' mapping between the bilinears of (4.7) and the 
conserved charges of the BRST formalism. Such a  suitable identification,
for our purposes,  is
$$
\begin{array}{lcl}
&&Q_{b} = A \tilde B \equiv Q \qquad Q_{d} = B \tilde A
\equiv \bar Q\;  \qquad Q_{b}^2 = Q_{d}^2 = 0 \nonumber\\
&& W = (A \tilde A - B \tilde B) \equiv H\; \qquad
- i Q_{g} = {\displaystyle \frac{A \tilde A + \mu B \tilde B} {1 + \mu}}
\equiv Y\;
\end{array} \eqno(6.2)
$$
and, in addition, the discrete symmetry transformations (5.1), (5.2) and
(5.3) do provide the realization of the Hodge duality $*$ operation of
differential geometry. We distinguish our realization of duality
(from the usual differential geometry Hodge $*$ duality) by denoting it by 
a separate and different $\star$ operation. 
To corroborate the above identifications beyond merely
an {\it algebraic equivalence},
we note that the conserved charges (ie $\dot Q_{b} = [ Q_b, H] = 0,
\dot Q_{d} = [Q_d, H] = 0$) generate the symmetry transformations 
$s_b$ and $s_d$ for the
Hamiltonian $H$. These transformations are encoded in the nilpotent
($ s_b^2 = s_d^2 = 0$)
operators $s_{b}$ and $s_{d}$. The explicit form of the transformations
generated by the conserved charges $Q_b = A \tilde B$ and $Q_d = B \tilde A$
for the {\it noncommutative} superoscillators
$A, \tilde A, B, \tilde B$, for the given superalgebra (4.6)
(ie $\nu = 0, \lambda = 1$ ), are
$$
\begin{array}{lcl}
&& s_{b} \tilde A = [ \tilde A, Q_{b} ] = - q^{-1} \tilde B + (1 - q^{-1})\;
\tilde A A \tilde B\; \nonumber\\
&&s_{d} \tilde A = [ \tilde A, Q_{d} ] = (1 - q)\; \tilde A B \tilde A\; \quad
 s_{b} A = [ A, Q_{b} ] = (q^{-1} - 1)\; A \tilde B A \nonumber\\
&& s_{d} A = [ A, Q_{d} ] = q B + (q - 1) \; B \tilde A A \nonumber\\
&&s_{b} \tilde B = \{ \tilde B, Q_{b} \} = 0\; \;\;\quad 
s_{d} \tilde B = \{ \tilde B, Q_{d} \} =  q \tilde A + (1 - q)\; \tilde B
B \tilde A \nonumber\\
&& s_{d} B = \{ B, Q_{d} \} = 0\; \quad
s_{b} B = \{ B, Q_{b} \} = A + (q^{-1} - 1) B \tilde B A.
\end{array} \eqno(6.3)
$$
It is very interesting to point out that the above transformations
are connected
to each-other by a general formula for the generic noncommutative
$q$-(super)oscillator $\Phi$ as
$$
\begin{array}{lcl}
\tilde s_{d}\; \Phi = \pm\; \star\; \tilde s_{b}\; \star\; \Phi
\end{array} \eqno(6.4)
$$
where $+$ sign on the r.h.s. is for $\Phi = B , \tilde B$ and $-$ sign 
on the r.h.s. is for $\Phi = A, \tilde A$ for {\it all} the cases of 
superalgebras (3.8), (4.4), (4.5) and (4.6).
The above signs are dictated by the general requirement of
a duality invariant theory (see, eg, [40] for details). We would like to lay
stress on the fact that the relationship in (6.4) is the analogue of
such type of relation that exists in the differential geometry as given
by equation (2.3). It will be
noted that, under all the above discrete transformations (ie (5.1)-(5.3)) 
corresponding to the $\star$ operation, the result of two successive 
$\star$ operation on the noncommutative $q$-superoscillators 
($A, \tilde A, B, \tilde B$) of $GL_{qp} (1|1)$ is:
$$
\begin{array}{lcl}
\star\; (\star \;A) = - A\; \qquad
\star\; (\star \;\tilde A) = - \tilde A\; \qquad
\star\; (\star \;B) = + B\; \qquad
\star\; (\star \;\tilde B) = + \tilde B\;
\end{array} \eqno(6.5)
$$
which decides the signatures present in (6.4). This observation should be
contrasted with the ($\pm$)-signs present in the relation
$\delta = \pm * d *$ between (co-)exterior derivatives $(\delta)d$ of
the differential geometry where these signs are dictated by the
dimensionality of the manifold on which these operators are defined.
The expressions for
$\tilde s_{b}$ and $\tilde s_{d}$ present in (6.4) are different for
various kinds of superalgebras. In fact, the relation (6.4) is valid for 
{\it all} the covariant superalgebras. For instance,
in the case of $\nu = 0, \lambda = 1$ which corresponds to $ pq = 1$, 
we have
$$
\begin{array}{lcl}
\tilde s_{d} = (-i q)^{(-1/2)}\; s_{d}\;\;\;\;
 \qquad \;\;\;\;\tilde s_{b} = (-i q)^{(+1/2)}\;
s_{b}.
\end{array} \eqno(6.6)
$$
It will be noted that the deformation parameters do not transform 
(cf (5.1)) in the above $\star$ operation corresponding to the covariant 
quantum superalgebra (4.8). For
the cases $\nu = 0, \lambda = pq$ and $ \nu = (pq)^{-1} (1 - pq),
\lambda = (pq)^{-1}$, we have 
$$
\begin{array}{lcl}
\tilde s_{d} = (+ i p)^{(+1/2)}\; s_{d}\;\;\;\; \qquad \;\;\;\tilde s_{b} 
= (-i q)^{(+1/2)}\; s_{b}.
\end{array} \eqno(6.7)
$$
Notice that, in the above transformations 
(cf (5.2) and (5.3)) corresponding to $\star$, the 
deformation parameters do transform and a close look at (6.6) and (6.7)
demonstrates that in the limit $ p = q^{-1}$, we do get back (6.6) 
from (6.7).

The sanctity and correctness of the relationship (6.4) can be checked
by computing the transformations generated by $Q_{b}$ and $Q_{d}$.
These transformations for the basic superoscillators 
$A, \tilde A, B, \tilde B$, for the algebra (3.8) and (4.4)
(ie for the case $ \nu = 0, \lambda = pq$), analogous to (6.3), are
$$
\begin{array}{lcl}
&& s_{b} \tilde A = [ \tilde A, Q_{b} ] = (1 - p^2 q)\;
\tilde A A \tilde B - p \tilde B\; \nonumber\\
&&s_{d} \tilde A = [ \tilde A, Q_{d} ] = (1 - q)\; \tilde A B \tilde A
\nonumber\\
&& s_{b} A = [ A, Q_{b} ] = (p - 1)\; A \tilde B A\; \quad
 s_{d} A = [ A, Q_{d} ] = q B + (pq^2 - 1) \; B \tilde A A \nonumber\\
&& s_{b} \tilde B = \{ \tilde B , Q_b\} = 0\; \;\;\;\qquad\;\;\;
s_{d} B = \{ B, Q_d \} = 0 \nonumber\\
&& s_{b} B = \{ B, Q_b \} =
A + (1 - q)\; B A \tilde B + (pq -1)\; A \tilde A A \nonumber\\
&& s_{d} \tilde B = \{ \tilde B, Q_d \} =
p^{-1} \tilde A + (1 - p^{-1})\; \tilde B B \tilde A
+ p^{-1}\; (pq -1)\; \tilde A A \tilde A.
\end{array} \eqno(6.8)
$$
The analogue of transformations (6.3) and (6.8) for the case of
superalgebra (3.8) and (4.5) that corresponds to $\nu = (pq)^{-1} (1 - pq),
\lambda = (pq)^{-1}$, are
$$
\begin{array}{lcl}
&& s_{b} \tilde A = [ \tilde A, Q{b} ] = (1 - q^{-1})\;
\tilde A A \tilde B - q^{-1} \tilde B \nonumber\\
&& s_{d} \tilde A = [ \tilde A, Q_{d} ] = (1 - p^{-1})\; \tilde A B \tilde A
\nonumber\\
&& s_{b} A = [ A, Q_{b} ] = (q^{-1} - 1)\; A \tilde B A\; \quad
 s_{d} A = [ A, Q_{d} ] = p^{-1} B +  p^{-1}\;(1- p) \; B \tilde A A
\nonumber\\
&& s_{b} \tilde B = \{ \tilde B , Q_b\} = 0\; \quad 
s_{d} \tilde B = \{ \tilde B, Q_d \} =
p^{-1} \tilde A + (1 - p^{-1})\; \tilde B B \tilde A\; \nonumber\\
&& s_{d} B = \{ B, Q_d \} = 0\; \quad
s_{b} B = \{ B, Q_b \} =
A + (1 - q)\; B A \tilde B. 
\end{array} \eqno(6.9)
$$
With the help of (6.5)--(6.7), it can be checked that the relationship
(6.4) is satisfied for all the transformations (6.3), (6.8) and (6.9). It is 
straightforward, in view of the relationship $\delta = \pm * d *$
between (co-)exterior derivatives of differential geometry, to claim that
the nilpotent transformations $\tilde s_{d}$ and $\tilde s_{b}$ are
dual to each-other. In other words, the discrete symmetry transformations
(5.1), (5.2) and (5.3)
for the covariant superalgebras (3.8), (4.4), (4.5) and (4.6) 
corresponding to the $\star$ operation (for the 
transformations connected with the $q$-superoscillators 
as well as the deformation parameters) are the
analogue of the Hodge duality $*$ operation of the differential geometry.

This relationship can also be established at the level of 
the conserved (ie $[H,  Q] = [H, \bar Q] = 0$) supercharges $Q$ and $\bar Q$
of the identification (4.7) that obey the algebra (4.8).
For instance, in the case of the unique superalgebra for $pq = 1$, 
we have the following relationship
between $Q$ and $\bar Q$ through $S$ and $\bar S$
$$
\begin{array}{lcl}
\bar S\; \Phi = \pm\; \star\; S\; \star\; \Phi\; \qquad\;\;\;
\bar S = (-iq)^{-1/2}\; \bar Q\; \qquad\;\;\; S = (-iq)^{1/2}\; Q
\end{array} \eqno(6.10)
$$
where the $+$ sign on the r.h.s. is for $\Phi = B , \tilde B$ and 
the $-$ sign on the r.h.s. is
for $\Phi = A, \tilde A$. Similar relations are valid for the cases
of the covariant quantum superalgebras when $pq \neq 1$
(ie (3.8) together with (4.4) and (4.5)). However, in those cases,
the $S$ and $\bar S$ are defined as
$$
\begin{array}{lcl}
\bar S = (+ip)^{+1/2}\; \bar Q\; \qquad\;\;\; S = (-iq)^{+1/2}\; Q.
\end{array} \eqno(6.11)
$$
It will be noted that the conservation of supercharges $Q, \bar Q$
can be recast in the language of the BRST-type transformations $s_{b}$
and $s_{d}$ which turn out to be the symmetry transformations for the
Hamiltonian $H$ as given below
$$
\begin{array}{lcl}
s_{b} H = [ H , Q_{b} ] = 0\; \qquad 
s_{d} H = [ H , Q_{d} ] = 0\; \qquad
\{ s_{b}, s_{d}\} H = [ H, \{Q_b, Q_d\} ] = 0.
\end{array} \eqno(6.12)
$$
This can also be re-expressed in the language of the de Rham
cohomological operators because $H \rightarrow \Delta$
is the Casimir operator for the whole algebra as
$ [ H, Q ] = [H, \bar Q] = 0 \rightarrow [H, Q_b] = [H, Q_{d}] = 0 \rightarrow
[\Delta, d] = [\Delta, \delta] = 0$. Hence in a single theoretical setting,
we have obtained a neat relationship among the BRST formalism, de Rham
cohomological operators and the covariant quantum superalgebras that
are constructed by the bilinears of the noncommutative $q$-superoscillators
for the supersymmetric quantum group $GL_{qp}(1|1)$.

To wrap up this section, we comment on (i) the transformations generated by
the operators $Y$ and $H$ of the identification in (6.2), and
(ii) the analogy between the 
variation of the degree of a form due to the operation of the cohomological
operators in differential geometry and the changes of the
ghost number of a state in the
quantum Hilbert space (QHS) due to the application of the conserved charges
in the framework of BRST formalism. Let us first concentrate on (i). The
corresponding transformations 
can be computed for all the algebraic relations (3.8), (4.4), (4.5)
and (4.6). For the sake of simplicity, however, let us focus only on the simple
case of
(4.6) (ie the case for $pq = 1$). For this algebra, $Y$ and $H$ generate
transformations that are encoded in the following commutators:
$$
\begin{array}{lcl}
&& [ \tilde A, Y ] = - (1/ 1 + \mu)\; \tilde A\; \;\;\qquad\;\;\;\;\;
\;\;\;\; \; [ A, Y ] = + (1/ 1 + \mu)\; A \nonumber\\
&& [ \tilde B, Y ] = - (\mu/ 1 + \mu)\; \tilde B\; \;\;\qquad\;\;\;\;\;
\;\;\; \; \; [ B, Y ] = + (\mu/ 1 + \mu)\; B \nonumber\\
&& [ \tilde A, H ] = - \tilde A\; \qquad
 [ A, H ] = + A\; \qquad
 [ \tilde B, H ] = - \tilde B\; \qquad
 [ B, H ] = + B.
\end{array} \eqno(6.13)
$$
On the face value, both the above transformations look the same modulo some
constant factors. However, a close look at the identification (6.2)
clarifies that there is a clear-cut distinction between the two because
of the presence of an $i$ factor in the expression for $Y  = - i Q_g$.
In fact, between the two transformations, one corresponds to a scale 
transformation and the other corresponds
to the gauge transformation. This is consistent with the ghost transformations
(generated by the conserved ghost charge) 
and the bosonic transformations (generated by a bosonic
charge that turns out to be the analogue of the Casimir operator) for
a duality invariant gauge theory 
described in the framework of BRST formalism (see, eg, [32-39] for details).
Now let us concentrate on (ii). The conserved charges $Q_{b}, Q_{d}$
and $H$ (which have been realized in terms of the noncommutative
$q$-superoscillators)
can be elevated to the operators in the QHS. Any arbitrary state
$|\Psi>_n$ with ghost number $n$ (ie $i Q_{g} |\Psi>_n = n |\Psi>_n$)
can be decomposed into a unique sum (in analogy with (2.2)) as
$$
\begin{array}{lcl}
|\Psi>_n = |\omega>_n + Q_b \; |\chi>_{(n - 1)} + Q_{d}\; |\theta>_{(n + 1)}
\end{array} \eqno(6.14)
$$
where $|\omega>_n$ is the harmonic state (ie $Q_b |\omega>_n 
= Q_d |\omega>_n = 0$)
and the nilpotent operators $Q_{(b)d}$ raise and lower the ghost
number of states $|\chi>_{(n-1)}$ and $|\theta>_{(n +1)}$ by one,
respectively. In more explicit and lucid language, it can be seen that
for the above state $|\Psi>_n$ with ghost number $n$, we have
$$
\begin{array}{lcl}
&& i Q_{g} Q_b \;|\Psi>_n =  (n + 1)\; Q_b |\Psi>_n \nonumber\\
&& i Q_g Q_d\; |\Psi>_n = (n - 1)\; Q_d |\Psi>_n \nonumber\\
&& i Q_g\; H\; |\Psi>_n = \;\;(n)\; H\; |\Psi>_n 
\end{array} \eqno(6.15)
$$
which shows that the ghost numbers for the states $Q_b |\Psi>_n, 
Q_{d} |\Psi>_n$ and $H |\Psi>_n$ (generated by the conserved charges
$Q_b, Q_d$ and $H$) are $(n + 1), (n -1)$ and $n$,
respectively. This also establishes the correctness of the identification
(6.2) of the bilinears of the $q$-superoscillators with the conserved
charges of the BRST formalism that, in turn, are connected with the
de Rham cohomological operators of differential geometry.\\

\noindent
{\bf 7 Conclusions}\\

\noindent
The central result of our present paper is to provide a realization of
the de Rham cohomological operators of differential geometry, Hodge
duality $*$ operation, Hodge decomposition theorem, etc, in the language
of (i) the noncommutative $q$-superoscillators, (ii) the covariant
quantum superalgebras of the bilinears in $q$-superoscillators, and
(iii) the discrete symmetry transformations on the $q$-superoscillators, etc, 
for a doubly deformed
supersymmetric quantum group $GL_{qp} (1|1)$. An interesting observation
in our present investigation is the fact that a {\it unique} covariant
supersymmetric quantum algebra emerges from a couple of consistent
$GL_{qp} (1|1)$ covariant superalgebras for the condition $pq = 1$
on the deformation parameters. Furthermore, the bilinears constructed from
the noncommutative $q$-superoscillators turn out to provide (i)
a realization of the $N = 2$ supersymmetric quantum mechanical algebra,
and (ii) a realization of an extended BRST algebra where there is
one-to-one mapping between a set of conserved charges 
$(Q_{b}, Q_{d}, W, -i Q_{g})$ of the BRST formalism and the
conserved supercharges $(Q, \bar Q)$,
the Hamiltonian $(H)$ and the Witten index $Y$ 
(that constitute a set ($Q, \bar Q, H, Y)$) of the $N = 2$
supersymmetric quantum mechanics, respectively. These charges, in turn, are
connected with the de Rham cohomological operators of the
differential geometry. Thus, our present
investigation sheds light on the inter-connections among the de Rham 
cohomological operators of differential geometry, an extended BRST
algebra (constituted by several conserved charges)
for a class of duality  invariant gauge theories
and the $N = 2$ supersymmetric quantum mechanical
algebra. All the above conserved charges and other operators
are expressed  in the language of noncommutative
$q$-superoscillators of a doubly deformed supersymmetric quantum group 
$GL_{qp} (1|1)$ in their various guises.

It is worth emphasizing that the Hodge duality $*$ operation of the
differential geometry appears in our discussion as a set of discrete
symmetry transformations under which a set of $GL_{qp}(1|1)$ covariant
superalgebras remain form invariant. This analogy and identification have been
established at two different and distinct levels of our discussion.
First, it turns out that the nilpotent (ie $ Q^2 = \bar Q^2 = 0$)
and conserved $(\dot Q = [ Q, H] = 0, \dot {\bar Q} 
= [\bar Q, H] = 0)$ supercharges $Q$ and 
$\bar Q$ (modulo some constant factors) are connected 
(cf (6.10)) with each-other
exactly in the same manner as the (co-)exterior derivatives $(\delta)d$
are related (ie $\delta = \pm * d *$) to each-other. Second, it is evident 
that the BRST-type transformations $s_{d}$ and $s_{b}$, generated by
$Q_{d} \equiv \bar Q$ and $Q_{b} \equiv Q$, are related
(cf (6.4)), modulo some
constant factors, in exactly the same way as the relationship
(ie $\delta = \pm * d *$) between (co-)exterior derivatives $(\delta)d$
of differential geometry defined on a manifold without a boundary.
At both level of identifications, the $\star$ operation turns out to be
equivalent to a set of discrete symmetry transformations (5.1), (5.2) and
(5.3), under which, a set of covariant quantum superalgebras 
(cf (3.8),(4.4),(4.5),(4.6)) remain form invariant.
The insight into such an identification comes basically from our experience
with the duality invariant gauge theories 
(that present a set of tractable field theoretical models
for the Hodge theory [32-39]) where
the discrete symmetry transformations, for the (co-)BRST invariant
Lagrangian densities, turn out to be the analogue of the Hodge duality
$*$ operation of differential geometry.

It is very much essential for our present algebraic discussions to,
ultimately, percolate down to the level of physical applications to
some interesting dynamical systems. In this context, it is interesting 
to pin-point that, in the language of differential geometry developed on
the super quantum hyperplane (see, eg, [41-43] and references therein for
details), the noncommutative $q$-(super)oscillators
can be identified with the Grassmannian as well as ordinary coordinates and
the corresponding derivatives. For instance, the set of bosonic oscillators
$(A, \tilde A)$ can be identified with an ordinary coordinate $x$
and the corresponding derivative $\partial/\partial x$, respectively. 
Similarly, the set of fermionic oscillators $(B, \tilde B)$ can be identified
with the Grassmannian coordinate $\theta$ and its corresponding derivative
$(\partial/\partial\theta)$, respectively, where $\theta^2 = 0$ and 
$(\partial/\partial\theta)^2 = 0$. These identifications, in turn, allow us 
to get a differential calculus on the super quantum hyperplane from
the covariant quantum algebra (3.8), (4.4), (4.5)
and (4.6) obeyed by the noncommutative  $q$-(super)oscillators. The
ensuing $GL_{qp} (1|1)$ covariant calculus will enable us to discuss
physical systems on the super quantum hyperplane with deformation
parameters $p$ and $q$.

It is interesting to  point out that a consistent formulation
of the dynamics on a noncommutative phase space has been developed where
the ordinary
Lorentz (rotational) invariance and the noncommutative
quantum group invariance are
maintained {\it together} for the quantum group $GL_{qp} (2)$ with
deformation parameters obeying $pq = 1$ [22,23]. Some of these ingredients
have also been exploited in the context of discussion of the Landau
problem in two dimensions where the Snyder's idea of noncommutativity
(reflected due to the presence of perpendicular constant magnetic
field for 2D electron system) and the noncommutativity due to quantum groups
$GL_{qp} (2)$ with $pq = 1$ are present {\it together} [23]. The
algebraic relations in the present paper and corresponding differential
calculus might turn out to be useful in the discussion of spinning
relativistic particle on a deformed super hyperplane.
In fact, this system has been discussed earlier [20] where only the on-shell
conditions (ie the equations of motion) have been exploited to obtain
the NC relations 
\footnote{We have assumed the relations $x_\mu x_\nu = x_\nu x_\mu, 
p_\mu p_\nu = p_\nu p_\mu, x_\mu p_\nu = q p_\nu x_\mu, 
\psi_\mu \psi_\nu + \psi_\nu \psi_\mu = 0$ in the phase space for the
spinning relativistic particle where $x_\mu$ and $p_\mu$ are the target
space (canonically conjugate) coordinates and momenta respectively
and the fermionic Lorentz vector $\psi_\mu$
stands for the ``spin'' degrees of freedom attached to the 
relativistic particle. In the above relations, the Lorentz invariance
is respected for any arbitrary ordering of $\mu$ and $\nu$. One of the 
highlights of this work is the $GL_{\surd q} (1|1)$ and $GL_{q}(2)$
invariance of the solutions for the equations of motion 
at any arbitrary value of
the parameter (ie time) of the evolution. The equations of motion are
derived from the Euler-Lagrange equations.} 
among the phase variables which play
a crucial role in the description of the dynamics on the noncommutative
($q$-deformed) phase space. However, a systematic
and consistent differential calculus has not been developed in [20] on
the super quantum hyperplane for such a discussion. We very strongly 
believe that our
present work will bolster up the derivation of
such a calculus on the super hyperplane. Similarly, some
supersymmetric field theoretic models can also be discussed in a 
systematic manner by exploiting the differential
calculus derived from the $q$-superoscillator algebra of our
present paper. These are the key issues 
that are under investigation
and our results will be reported elsewhere.\\

\noindent
{\bf Acknowledgements}\\

\noindent
It is a pleasure to thank the referees for their very
constructive and clarifying comments. 
This work was initiated at the AS-ICTP, Trieste, Italy. Fruitful discussions
with A. Klemm, K. S. Narain and G. Thompson are gratefully acknowledged. 
This paper is inspired by a set of lectures given by A. Klemm in the
``School on Mathematics in String and Field Theory'' (2-12 June 2003)
held at the AS-ICTP, Trieste, Italy. The warm hospitality extended by
the HEP group to me at AS-ICTP, Trieste, Italy is gratefully 
acknowledged too.\\

\begin{center}

{\bf Appendix}

\end{center}

\noindent
To establish the uniqueness of the algebras in (3.8), (4.4), (4.5)
and (4.6), we show that, given a trilinear 
combination of the (super)oscillators, we
can arrange all the tilde oscillators to the left in two different ways
due to the requirement of the associativity condition.
In the following, the 
{\it pair} of the (super)oscillators that are exchanged {\it first},
due to the covariant algebras given in section 3,
are kept within the round brackets (cf. l.h.s. of (A.1),(A.2),etc., below). 
It will be noted that, on the r.h.s, there is no such reordering because
the trilinears on the r.h.s. are {\it unique} in the sense that all the
tilde oscillators have been brought to the left due to the algebras of 
section 3 for comparison. In fact, it is the comparison on the r.h.s. of
the reordering
\footnote{This requirement is the very essence of the validity of the
graded Yang-Baxter equations.} that determines
the exact values for the parameters $\lambda$ and $\nu$. To corroborate the
above assertion, we take here a set of four trilinears of the (super)oscillators
and show that, the requirement of the associativity condition, yields
the same values for $\lambda$ and $\nu$ for all the members of this set. 
In fact, the rearrangements
of the four members of the trilinears of the (super)oscillators
$$
\begin{array}{lcl}
A\; (\tilde A \tilde B) &=& 
\Bigl [\;\Bigl ( {\displaystyle \frac{\lambda - \nu}{pq}}\;
\Bigr )\;\Bigr ]
\; \tilde B +
\Bigl [\;\Bigl ({\displaystyle \frac{\lambda (\lambda - \nu)}{pq}}\;
\Bigr )\;\Bigr ]\;
\tilde B \tilde A A \nonumber\\
(A \tilde A)\; \tilde B &=& \Bigl [ (1 + \nu)
\Bigr ]\; \tilde B + \Bigl [\;
{\displaystyle \frac{\lambda (\lambda - \nu)}{pq}}
 + \nu \Bigl (\; \lambda -
{\displaystyle \frac{\lambda - \nu}{pq}}\;\Bigr )\; \Bigr ]\;
\tilde B \tilde A A 
\end{array}\eqno (A.1)
$$
$$
\begin{array}{lcl}
B\; (\tilde A \tilde B) &=& 
\Bigl [\;{\displaystyle \frac{1}{p}\;
\Bigl ( 1 + \lambda - \frac{\lambda - \nu}{pq}} \Bigr ) \;\Bigr ]
\; \tilde A + 
\Bigl [\;{\displaystyle \frac{\lambda}{p}\;\Bigl ( \lambda -
\frac{\lambda - \nu}{pq}}\;\Bigr )\; \Bigr ]\;
\tilde A \tilde A A \nonumber\\
&+& 
\Bigl [\;{\displaystyle \frac{\lambda}{p}\;\Bigl ( \lambda -
\frac{\lambda - \nu}{pq}}\;\bigr ) \Bigr ]\;\tilde A \tilde B B \nonumber\\
(B \tilde A)\; \tilde B &=& \Bigl [\;\Bigl (
 {\displaystyle \frac{\lambda - \nu}{p}} \Bigr )\;
\Bigr ]\; \tilde A + \Bigl [\;
\Bigl ( {\displaystyle \frac{\lambda - \nu}{p} \Bigr )
\Bigl ( \lambda - \frac{\lambda - \nu}{pq}} \Bigr ) \;\Bigr ]\;
\tilde A \tilde A A \nonumber\\
&+& \Bigl [\; \Bigl ( {\displaystyle \frac{\lambda - \nu}{p} \Bigr )
\Bigl ( \nu - \frac{\lambda - \nu}{pq}} \Bigr )\; \Bigr ]\;
\tilde A \tilde B B
\end{array}\eqno (A.2)
$$
$$
\begin{array}{lcl}
A\; (B \tilde B) &=& 
\Bigl [\;1 + 
\Bigl ( \lambda - {\displaystyle \frac{\lambda - \nu}{pq}} \Bigr ) \;\Bigr ]
\; A +
\Bigl [\;\lambda \;\Bigl ( \lambda -
{\displaystyle \frac{\lambda - \nu}{pq}}\;\Bigr )\; \Bigr ]\;
\tilde A A A \nonumber\\
&+& \Bigl [\;\nu \;\Bigl ( \lambda -
{\displaystyle \frac{\lambda - \nu}{pq}}\;\Bigr ) +
\Bigl (\lambda - \nu \Bigr) \Bigl ( \nu -
{\displaystyle \frac{\lambda - \nu}{pq}}\; \Bigr )\;
\Bigr ] \; \tilde B B A
\nonumber\\
(A B)\; \tilde B &=& \Bigl [ \Bigl (\lambda - \nu \Bigr )\; \Bigr ]\; A
+ \Bigl [\;
\Bigl ( \lambda - \nu \Bigr )
\Bigl ( \lambda - {\displaystyle \frac{\lambda - \nu}{pq}} \Bigr )\; \Bigr ]\;
\tilde A A A \nonumber\\
&+& \Bigl [\; \Bigl (\lambda - \nu \Bigr )
\Bigl ( \nu - 
{\displaystyle \frac{\lambda - \nu}{pq}} \Bigr )\; \Bigr ]\;
\tilde B B A
\end{array}\eqno (A.3)
$$
$$
\begin{array}{lcl}
B\; (A \tilde A) &=& \Bigl [\; \Bigl (1 + \nu \Bigr )\;\Bigr ]\; B
+ \Bigl [\; {\displaystyle \frac{\lambda (\lambda - \nu )}{pq}}
+ \nu \Bigl ( \lambda - {\displaystyle \frac{\lambda - \nu}{pq}}
\Bigr )\; \Bigr ]\; \tilde A A B \nonumber\\
(B A)\;\tilde A &=& \Bigl [\;
\Bigl ( {\displaystyle \frac{\lambda - \nu}{pq}} \Bigr )\; \Bigr ]\; B
+ \Bigl [\;\Bigl ( {\displaystyle \frac{\lambda (\lambda - \nu)}{pq}}
\Bigr )\; \Bigr ]\; \tilde A A B 
\end{array}\eqno (A.4)
$$
lead to a unique set of relations between $\lambda$ and $\nu$
$$
\begin{array}{lcl}
\lambda = p q + (pq + 1) \nu \;\;\;\qquad\;\;\;
\nu \Bigl (\; \lambda - {\displaystyle \frac{\lambda - \nu}{pq}}\; \Bigr )
= 0
\end{array}\eqno (A.5)
$$
when the r.h.s. of the above equations (i.e. (A.1)-(A.4)) are matched
with each-other.
It is straightforward to check that the above relations lead to
the set of values of $\lambda$ and $\nu$ as quoted in (4.3). This discussion
demonstrates the uniqueness of the algebras (3.8), (4.4) and (4.5). Of course,
the algebra (4.6) is a special case of the above algebras when $pq = 1$.

\end{document}